\documentclass[final,numberedheadings]{aipproc}
\layoutstyle{6x9}

\usepackage{latexsym,verbatim,amsmath,graphicx}

\begin{document}
\title{A quantum cosmological model in Ho\v{r}ava-Lifshitz gravity}


\author{O. Obreg\'{o}n}
{address={Departamento de F\'{\i}sica, Divisi\'{o}n de Ciencias e Ingenier\'{\i}as, Campus Le\'{o}n, Universidad de Guanajuato, A.P. E-143, C.P. 37150, Le\'{o}n, Guanajuato, M\'{e}xico.}}

\author{J. A. Preciado}
{address={Departamento de F\'{\i}sica, Divisi\'{o}n de Ciencias e Ingenier\'{\i}as, Campus Le\'{o}n, Universidad de Guanajuato, A.P. E-143, C.P. 37150, Le\'{o}n, Guanajuato, M\'{e}xico.}}

\begin{abstract}
    A Wheeler-DeWitt equation for the Kantowski-Sachs model is derived within the framework of the minimal quantum gravity theory proposed by Ho\v{r}ava. We study the solution to this equation in the ultraviolet limit for the specific case where the $\lambda$ parameter of the theory takes its relativistic value $\lambda = 1$. It is observed that the minisuperspace variables switch their role compared with their usual infrared (General Relativity) behavior.
\end{abstract}

\maketitle

\section{Introduction}

Recently Petr Ho\v{r}ava proposed an interesting approach to quantum gravity \cite{Horava1,Horava2}. The central idea of his theory is to combine gravity with the concept of anisotropic scaling between space and time, motivated by the recent developments in the study of condensed matter systems. Based on this principle, higher-derivative correction terms may be added to the standard Einstein-Hilbert action such as different powers of the spatial curvature. This improves the ultraviolet (UV) behavior of the graviton propagator and makes the theory power counting renormalizable, but at the cost of giving up to Lorentz invariance as a fundamental symmetry. Instead of that, it is expected to emerge as an accidental symmetry in the infrared (IR) regime where general covariance must be restored.

In the minimal version of the theory the nature of the modifications is governed by the gravitational analog of the ``detailed balance'' principle, frequently used in the study of the dynamics of nonequilibrium systems, and the so called ``projectability condition'' which restricts the lapse function to be a function of time only. However, it is possible to construct a generalization by relaxing any or both of these conditions. This has lead to the projectable and non-projectable versions of the theory \cite{Visser1,Blas1}.

It has been claimed that Ho\v{r}ava's proposal and its extensions suffer from several issues related to a badly behaved scalar mode \cite{Niz,Blas2}. However, it has many desirable features and seems worth exploring it. Many aspects of the theory have been discussed in the literature, particularly, cosmological and black hole solutions have been obtained \cite{Calcagni,Pope}. Many issues of cosmology arising from it have also been analyzed (see for example \cite{Mukohyama}). However, no model that provides information about the quantum properties of this theory seems to have been considered up to now.

For this purpose in this work we consider the Kantowski-Sachs (KS) universe which is one of the simplest anisotropic models. A Wheeler-DeWitt (WDW) equation for this model is derived in the context of the minimal theory proposed by Ho\v{r}ava. We study the solution to this equation in the UV regime for the specific case in which the $\lambda$ parameter of the theory takes its relativistic value $\lambda = 1$ by constructing a gaussian-weighted wave packet. It is observed that the minisuperspace variables switch their role in this limit compared to their usual General Relativity (GR) behavior.

In section \ref{Theory} we begin by briefly presenting the general features of the gravity models with anisotropic scaling. In section \ref{KSGR} we review the WDW equation for the KS model in the context of GR. In section \ref{KSHL} a WDW equation for this model is derived within the framework of the minimal theory proposed by Ho\v{r}ava. Particularly, the UV regime is analyzed and the behavior of the resulting wave packet is compared with the already known results in the context of GR. Finally Section \ref{Conclusions} is devoted for discussion and conclusions.

\section{The Theory}\label{Theory}
In field theories with anisotropic scaling the degree of anisotropy between the space and time coordinates is characterized by a dynamical critical exponent $z$ such that
\begin{equation}\label{AS:z}
  \mathbf{x}\rightarrow b\mathbf{x},\qquad t\rightarrow b^{z}t.
\end{equation}
The gauge symmetries are a reduced set of spacetime diffeomorphisms, denoted by $\text{Diff}_{\mathcal{F}}(\mathcal{M})$, that preserve a preferred foliation of spacetime by fixed time slices generated by the infinitesimal transformations $x^{i} \rightarrow \tilde{x}^{i}(t,x^{i})$ and $t \rightarrow \tilde{t}(t)$. In this context it is convenient to consider the ADM decomposition of spacetime and to construct the action in terms of the spatial metric $g_{ij}$, the shift vector $N_{i}$ and the lapse function $N$. With these ingredients the most general action of this class of gravity models takes the form
\begin{equation}\label{Action:AS}
  S = \frac{2}{\kappa^{2}} \int dtd^{D}x \sqrt{g}N \left\{K_{ij}K^{ij} - \lambda K^{2} + \mathcal{V} \right\},
\end{equation}
where $\kappa$ and $\lambda$ are coupling constants and $K_{ij}$ is the extrinsic curvature tensor. The first two terms represent the most general kinetic term invariant under $\text{Diff}_{\mathcal{F}}(\mathcal{M})$ and $\mathcal{V}$ is an arbitrary potential term built on $g_{ij}$ and its spatial derivatives, compatible with $\text{Diff}_{\mathcal{F}}(\mathcal{M})$ and the desired value of $z$. Power-counting arguments require $z = 3$ in $3+1$ dimensions. As a first attempt, Ho\v{r}ava restricted $\mathcal{V}$ to satisfy the so called ``detailed balance'' and ``projectability'' conditions. The action of this theory under these considerations in $3+1$ dimensions is given by
\begin{align}\label{HoravaAction}
  \nonumber S =\int dtd^{3}x\sqrt{g}N &\left\{\frac{2}{\kappa^{2}}(K_{ij}K^{ij} - \lambda K^{2}) - \frac{\kappa^{2}}{2w^{4}}C_{ij}C^{ij} + \frac{\kappa^{2}\mu}{2w^{2}}\varepsilon^{ijk}R_{il}\nabla_{j}R^{l}_{k} - \frac{\kappa^{2}\mu^{2}}{8}R_{ij}R^{ij}\right.
  \\ & \left. + \frac{\kappa^{2}\mu^{2}}{8(1-3\lambda)} \left(\frac{1-4\lambda}{4}R^{2}+\Lambda_{W}R-3\Lambda_{W}^{2}\right)\right\},
\end{align}
where $\mu$, $w$ and $\Lambda_{W}$ are constant parameters and $C_{ij}$ is the Cotton tensor.
Comparing with the Einstein-Hilbert action in the ADM formalism, the speed of light, Newton's constant and the cosmological constant emerge as
\begin{equation}\label{cGL}
  c = \frac{\kappa^{2}\mu}{4} \sqrt{\frac{\Lambda_{W}}{1-3\lambda}}, \qquad G_{N} = \frac{\kappa^{2}}{32\pi c}, \qquad \Lambda = \frac{3}{2} \Lambda_{W}.
\end{equation}
Furthermore, the requirement that this action be equivalent to the standard Eintein-Hilbert action in the IR limit requires that $\lambda$ takes its relativistic value $\lambda = 1$.

\section{Kantowski-Sachs model in General Relativity}\label{KSGR}
Let's start by reviewing the well-known quantum solutions to the KS universe in the context of GR. The Misner parametrization for this metric is \cite{Misner1972}
\begin{equation}\label{KSmetric}
ds^2 = -N(t)^{2}dt^{2} + e^{2\sqrt{3}\beta(t)}dr^{2} + e^{-2\sqrt{3}\beta(t) - 2\sqrt{3}\Omega(t)} (d\theta^{2} + \sin^{2}\theta d\varphi^{2}).
\end{equation}
Following the usual procedure of quantum cosmology the GR Hamiltonian density for this model is given by
\begin{equation}\label{Hamiltonian:GR}
\mathcal{H}_{EH} = N\mathcal{H}_{\bot} = \frac{N e^{\sqrt{3}\beta + 2\sqrt{3}\Omega}}{24} \Big[\Pi_{\beta}^{2} - \Pi_{\Omega}^{2} - 48e^{-2\sqrt{3}\Omega}\left(1 - \Lambda e^{-2\sqrt{3}\beta - 2\sqrt{3}\Omega} \right) \Big].
\end{equation}
Variation with respect to the Lagrange  multiplier $N$ yields the Hamiltonian constraint $\mathcal{H}_{\bot} \approx 0$. Then we may proceed to quantize the model using the canonical quantization recipe given by Dirac promoting the minisuperspace variables to operators and implementing this constraint as a restriction on the physically allowed wave functions $\mathcal{H}_{\bot}\psi = 0$. With a particular factor ordering and through the usual representation for the momenta $\Pi_{\beta}=-i \frac{\partial}{\partial{\beta}}$ and $\Pi_{\Omega}=-i \frac{\partial}{\partial{\Omega}}$ we get the WDW equation
\begin{equation}\label{WDW:GR}
\left[ -\frac{\partial^{2}}{\partial\Omega^{2}} + \frac{\partial^{2}}{\partial\beta^{2}} + 48e^{-2\sqrt{3}\Omega}\left(1 - \Lambda e^{-2\sqrt{3}\beta - 2\sqrt{3}\Omega} \right)\right] \psi(\beta,\Omega)=0.
\end{equation}
The solution to this equation with $\Lambda=0$ was found by Misner \cite{Misner1972} and is given by
\begin{equation}\label{WDW:GR:sol}
\psi_{\nu}^{\pm}(\beta,\Omega) = e^{\pm i \nu \sqrt{3} \beta} K_{i\nu}(4e^{-\sqrt{3}\Omega}),
\end{equation}
where $\nu$ is a separation constant and $K_{i\nu}$ are the modified Bessel functions of imaginary order.

\section{Kantowski-Sachs Model in Ho\v{r}ava gravity}\label{KSHL}
\begin{figure}[b!]
    \resizebox{.47\textwidth}{!}
    {\includegraphics{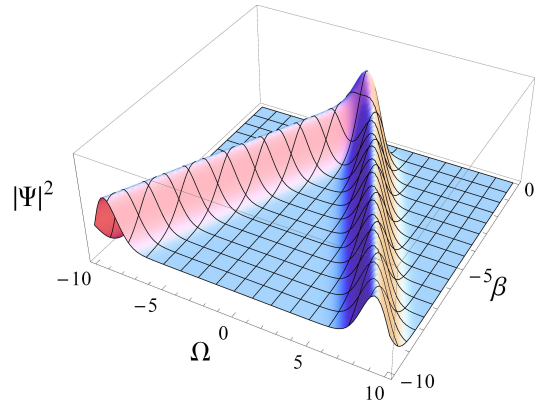}}
    \caption{Variation of $|\Psi|^{2}$ with respect to $\Omega$ and $\beta$ for $\lambda = 1$ (UV limit).}
    \label{WPHorava}
\end{figure}

\begin{figure}[t!]
    \resizebox{.47\textwidth}{!}
    {\includegraphics{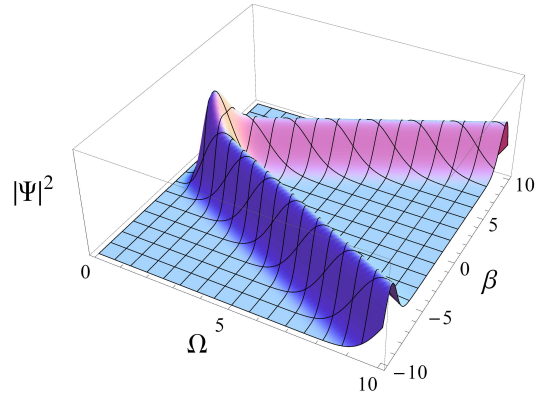}}
    \caption{Variation of $|\Psi|^{2}$ with respect to $\Omega$ and $\beta$ in GR (IR limit).}
    \label{WPEinstein}
\end{figure}

Now we proceed similarly to study the KS model within the framework of the minimal theory proposed by Ho\v{r}ava. By inserting the metric ansatz \eqref{KSmetric} into the action \eqref{HoravaAction} we may derive the corresponding Hamiltonian density for this model. Then using the canonical quantization recipe with a particular factor ordering, we get the WDW equation
\begin{align}\label{WDW:HL}
  \nonumber \bigg\{ & \frac{1}{2}(\lambda-3)\frac{\partial^{2}}{\partial\Omega^{2}} - 2(\lambda-1)\frac{\partial}{\partial\beta}\frac{\partial}{\partial\Omega}  +(2\lambda-1)\frac{\partial^{2}}{\partial\beta^{2}}
  \\ &- 3\mu^{2}\Lambda_{W}e^{-2\sqrt{3}\Omega} \bigg[2 - 3\Lambda_{W}e^{-2\sqrt{3}\beta-2\sqrt{3}\Omega} + \frac{(2\lambda-1)}{\Lambda_{W}}e^{2\sqrt{3}\beta+2\sqrt{3}\Omega}\bigg]\bigg\}\psi(\beta,\Omega)=0.
\end{align}
This equation considerably differs from the usual WDW equation \eqref{WDW:GR} but notice that for this particular model with $\lambda=1$ and making use of the expressions \eqref{cGL} in units such that $c=1$ and $\tfrac{1}{16\pi G} = 1$, it reduces to \eqref{WDW:GR} with an additional potential term coming from the higher-order terms in \eqref{HoravaAction} and which is responsible for the UV behavior of the model. Since an analytical solution for this equation is difficult to be found we consider the UV limit where the last term in the potential domains. The reduced WDW equation is
\begin{equation}\label{WDW:HL:UV}
   \bigg[ \frac{1}{2}(\lambda-3)\frac{\partial^{2}}{\partial\Omega^{2}} - 2(\lambda-1)\frac{\partial}{\partial\beta}\frac{\partial}{\partial\Omega} + (2\lambda-1)\frac{\partial^{2}}{\partial\beta^{2}} - 3\mu^{2}(2\lambda-1)e^{2\sqrt{3}\beta}\bigg]\psi(\beta,\Omega) = 0,
\end{equation}
whose solution for $\lambda = 1$ is given by
\begin{equation}\label{WDW:HL:UV:sol}
  \psi_{\nu}(\beta,\Omega)=e^{i\nu\sqrt{3}\Omega} K_{i\nu} \left(\mu e^{\sqrt{3}\beta}\right).
\end{equation}
In order to see the consequences of the UV corrections let us consider a wave packet weighted by a Gaussian centered in $\nu = \bar\nu$ and with standard deviation $\sigma$
\begin{equation}\label{WavePacket}
  \Psi(\beta,\Omega) = \int_{-\infty}^{\infty} e^{-\frac{1}{2\sigma^{2}}(\nu - \bar\nu)^{2}} \psi_{\nu}(\beta,\Omega)d\nu.
\end{equation}
The integral is performed numerically for the specific values $\sigma^{2} = 1/3$ and $\bar\nu=1.3$ with $\mu = 4$. Not only are we interested to see the influence of the second order spatial curvature terms, but also we want to study the behavior of the probability depending on the values of the $\beta$ and $\Omega$ variables. Figure \ref{WPHorava} shows the variation of the square of the wave packet magnitude $|\Psi|^{2}$ as a function of the minisuperspace variables $\beta$ and $\Omega$ for the case $\lambda=1$. It can be seen that in this case there is only one preferred state of the universe around $\beta = -1.5$ and $\Omega = 0$. Figure \ref{WPEinstein} shows the already known IR behavior in GR \cite{Obregon} described by the solution \eqref{WDW:GR:sol}. It can be seen that both graphs are quite similar, however, interestingly enough, the minisuperspace variables switch their role between these two limits. This, in fact, can be easily observed by comparing the solution \ref{WDW:GR:sol} with \ref{WDW:HL:UV:sol} under the replacements $\pm\beta \rightarrow \rightarrow \Omega$ and $-\Omega \rightarrow \beta$.

\section{Conclusions and Discussion}\label{Conclusions}
In this work we have studied the KS universe into the minimal version of Ho\v{r}ava gravity. A WDW equation was derived in this framework and solutions for the UV limit of this model for the case $\lambda = 1$ were analytically obtained. The analysis of the resulting solutions was performed by constructing a wave packet weighted by a gaussian amplitude. It was observed that compared to the IR (GR) limit, the minisuperspace variables switch their role in the UV regime for this particular case. This behavior may be of great interest into the quantum cosmology treatment of anisotropic models where the $\Omega$ variable is usually related to an intrinsic time. In this context it would be really interesting to extract dynamical information of the model in order to see the effects of the additional terms in the classical arena.

\begin{theacknowledgments}
  J.A.P acknowledges support from the CONACYT-M\'exico Scholarship Programme. This work was partially supported by CONACYT Grant 135023, PROMEP, and UG Projects.
\end{theacknowledgments}

\bibliographystyle{aipproc}   

\end{document}